\documentclass[preprintnumbers,amsmath,amssymb]{revtex4}
\usepackage{graphicx}
\usepackage{dcolumn}% Align table columns on decimal point
\usepackage{bm}% bold math
\usepackage{graphicx}
\usepackage{epsfig}
\usepackage[usenames]{color}
\usepackage{tikz}
\usetikzlibrary{shapes}

\usepackage{setspace}
\usepackage[mathcal]{eucal}
\usepackage{subfig}
\usepackage{amsthm}
\usepackage{float}
\usepackage[utf8]{inputenc}  %for apostrophe

\newcommand*\oiint{%
	\tikz \node[draw, ellipse, inner xsep=-3pt, inner ysep=-9pt] {$\displaystyle\iint_{\vec{\mathcal S}}$};%
	}
	
	\newcommand*\oiiint{%
	\tikz \node[draw, ellipse, inner xsep=-4pt, inner ysep=-9pt] {$\displaystyle\iiint_{\mathcal V}$};%
	}
\newcommand{\df}[2]{\displaystyle\frac{#1}{#2}}
\newcommand{\tf}[2]{\textstyle\frac{#1}{#2}}

\newcommand{\PDD}[2]{\df{\partial #1}{\partial #2}}

\newcommand{\eps}{\varepsilon}
\newcommand{\B}[1]{\mbox{\boldmath $ #1 $} }

\newcommand{\be}{\begin{eqnarray}}
\newcommand{\en}{\end{eqnarray}}

\begin{document}

\title{Micro cavitation bubbles on the movement of an experimental submarine\\ 
\it{Theory and Experiments}}

\author{Stefan C. Mancas}
\email{mancass@erau.edu}
\affiliation{Department of Mathematics, Embry-Riddle Aeronautical University,\\ Daytona-Beach, FL. 32114-3900, U.S.A.}
\author{Shahrdad G. Sajjadi}
\email{sajja8b5@erau.edu}

\affiliation{Department of Mathematics, Embry-Riddle Aeronautical University,\\ Daytona Beach, FL. 32114-3900, U.S.A. and \\
Trinity College, University of Cambridge, U.K}

\author{Asalie Anderson, Derek Hoffman}
%\email{hcr@ipicyt.edu.mx}
\affiliation{Department of Mathematics, Embry-Riddle Aeronautical University,\\ Daytona-Beach, FL. 32114-3900, U.S.A}

%\pacs{02.30.Hq, 02.30.Ik, 02.30.Gp}
%ordinary diff eqs, integrable systems, special functions

\begin{abstract}
To further understand their nature, micro cavitation bubbles were systematically diffused around the exterior of a test body (tube) fully submerged in a water tank. The primary purpose was to assess the feasibility of applying micro cavitation as a means of depth control for underwater vehicles, mainly but not limited to submarines. Ideally, the results would indicate the use of micro cavitation as a more efficient alternative to underwater vehicle depth control than the conventional ballast tank method. The current approach utilizes the Archimedes\rq{} principle of buoyancy to alter the density of the object affected, making it less than, or greater than the density of the surrounding fluid. However, this process is too slow for underwater vehicles to react to sudden obstacles inherent in their environment. Rather than altering its internal density, this experiment aimed to investigate the response that would occur if the density of its environment was manipulated instead. In theory, and in a hydrostatic fluid, diffusing micro air bubbles from the top surface of the submarine would dilute the column of water above it with air cavities, thus lowering the density of the water. The resulting pressure differential would then cause the submarine to gain buoyancy. Conversely, diffusing micro cavities underneath the submarine would reduce its buoyancy. By this reasoning, the greater the rate of air bubbles diffused, the greater the effects would be. The independent variable in this experiment was the flow rate of diffused air, while the variable being affected was the amount of change in the buoyancy of the submarine. The results of the experiment indicated an increase in buoyancy regardless of where the micro cavities were diffused. In fact, no correlation was found between the rate at which air bubbles were diffused and the vehicle's buoyancy. Instead, it depended on the amount of air bubbles forming on the diffuser at the time, and the size of bubbles. The paper is organized as follows: in section I we introduce the experiment.  In section II we present a mathematical model of a vacuous cavitation bubble where we will show how we find the radius of the bubble and its collapsing time. Also, we will calculate the time of expansion of a submarine mine prior to collapsing.  In section III we will back up the theory using an experiment performed in a water tank, and we conclude with results and comments. 

\end{abstract}
\maketitle

\section{Introduction}

This research attempts to analyze the feasibility of micro cavitation bubbles as a method of depth control for underwater vehicles. 

Underwater vessels rely mostly on ballast tanks pressurization to control their depth. Compressed air and water are allowed to alternately fill the tanks, varying the overall density of the vehicle. Filling the ballasts with compressed air decreases the vehicle's density and allows it to rise, while filling them with water incurs the opposite \cite{Pearce}. Though this method of depth control has been widely accepted and constantly modified, the process of diving and surfacing are often slow due to the rate at which water and air can be pumped into the ballast tanks. Conversely, underwater vehicles often encounter harsh scenarios in their environment with a multitude of random obstacles. Thus, in circumstances that demand rapid diving or surfacing of an undersea vessel, the method of altering the vessel's internal density may prove inadequate.

In order to satisfy this demand, this research aimed to enable underwater vehicles to rapidly dive or rise by altering the surrounding fluid density rather than internal vehicle density. This was  accomplished through systemic cavitation, or the introduction of small air bubbles near the surface boundary of an underwater vessel. 

This experiment aimed to prove or disprove two aspects of a hypothesis: 1) when micro cavitation bubbles were diffused from the top of the submarine, they would lower the density of the column of water above the submarine, thereby increasing its buoyancy. Conversely, when micro cavitation bubbles were diffused from the bottom of the submarine, they would decrease the density of the water directly adjacent to the bottom surface of the submarine, thus causing it to lose buoyancy and sink;  2) the amount of change in buoyancy for both instances would be dependent on the amount of air bubbles diffused into the surrounding water. The more air diffused, the greater the change in buoyancy, and vice versa.

Currently, only limited research has been conducted on the dynamic interactions between micro cavitation bubbles and surrounding fluid, underwater objects, and other micro cavitation bubbles. Furthermore, no data has been acquired on how the bubbles respond to subsurface waves, which would be a frequent occurrence in real world scenarios. Therefore, an experimental approach was deemed an efficient and productive method to initiate this research. Because little was known about how the bubbles would affect the motion of the undersea vessel, the project also aimed to uncover any potential issues concerning stability and navigational control of the vessel when applying this technology.

The first aspect of the hypothesis could be justified by the following derivations. Consider the case when a body was fully submerged under water as shown in Fig. \ref{figure 11} below.

\begin{figure}[H]
\centering
\includegraphics[width=0.5\textwidth]{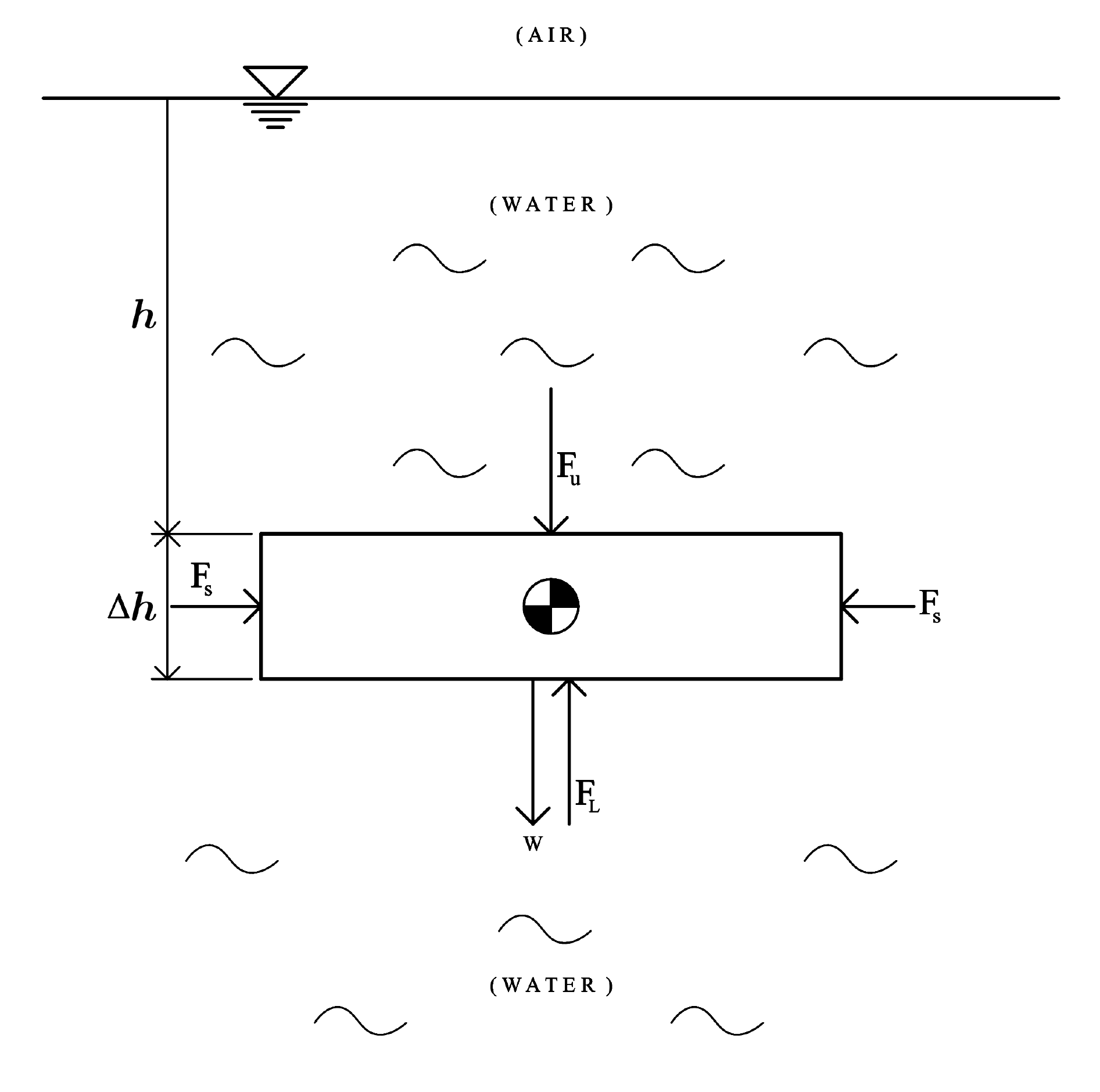}%this is figure 1.1%
\caption{\small{A cuboid fully submerged under water at depth $h$ from the surface. $F_U, F_L$, and $F_S$ are the equivalent forces acting on the cuboid due to water pressure. $W$ is the weight of the cuboid.}}
\label{figure 11}
\end{figure}

Assume that the body is a cuboid in a hydrostatic fluid, and that the fluid has uniform density. For an area $A$ the pressure exerted by the water on the cuboid can be expressed as shown in \eqref{s1} \cite{Yon}
 \begin{equation}\label{s1}
 p=\frac{dF_{\bot}}{d A}
 \end{equation}	
 				
where $p$ is the pressure exerted by the water on the cuboid, in $Pa$, $F_{\bot}$ is the equivalent force due to the water pressure, in $N$, and $A$  is the surface area exposed to the water pressure, in $m^2$.
						
Since the water has uniform density and assuming that the cuboid is leveled, the pressure acting on the top and bottom surfaces of the cuboid is also uniform. As a result, \eqref{s1} can be rearranged into \eqref{s3}. Hence, the equivalent forces acting on the top $F_U$ and bottom $F_L$ sides of the cuboid can be expressed as shown in \eqref{s4},\eqref{s5} respectively. The water pressure acting on the sides of the cuboid varies with depth, but since the cuboid is leveled, the pressure acting on the opposite sides of the cuboid balances each other resulting in equilibrium. 
  \begin{equation}\label{s3}
 F_{\bot}=pA
 \end{equation}
  \begin{equation}\label{s4}
F_U=p_UA
 \end{equation}
  \begin{equation}\label{s5}
F_L=p_LA
 \end{equation}				
					
In a hydrostatic fluid, the absolute pressure acting on the top surface of the cuboid at depth $h$ can be expressed as shown in \eqref{s6}\cite{Cen} 		
 \begin{equation}\label{s6}
p_U=p_{atm}+\rho_{W} g h
 \end{equation}			
 
where $p_U$ is the pressure acting on the top side of the cuboid, in $Pa$, $p_{atm}$ is the atmospheric pressure at the surface of water $=101,325 ~Pa$,
$\rho_{W}$  is the equivalent force due to the water pressure $=1000~kg/m^3$, $ g$ is gravitational acceleration $=9.81~m/sec^2$, and $h$  is depth of cuboid,  in $m$,  as illustrated in Fig. \ref{figure 11}.
		
Similarly, the absolute pressure acting on the bottom side of the cuboid can be expressed as
    \begin{equation}\label{s7}
p_L=p_{atm}+\rho_{W} g h+\rho_{W} g \Delta h
 \end{equation}
where $p_L$ = the water pressure acting on the bottom side of the cuboid,  in $Pa$,		
and $\Delta h$ is the thickness of the cuboid, in $m$, as illustrated in Fig. \ref{figure 11}.
 
By definition, the buoyant force acting on a body in a submerged fluid is the difference in the forces acting on the top and bottom sides of the body due to the pressure difference of the fluid \cite{Cen}, and it is shown in \eqref{s8}. Assuming that the cuboid is constructed such that its weight is equal to the buoyant force it experiences in still water, it should therefore be in equilibrium in all directions. 
 
\begin{equation}\label{s8}
F_{B_{init}}=F_L-F_U%p_LA-p_U A
= (p_{atm}+\rho_W g h +\rho_W g \Delta h)A-(p_{atm}+\rho_W gh)A=\rho_{W} g \Delta h A>0
 \end{equation}	
		
where $F_{B_{init}}$ is the initial  buoyant force acting on the cuboid in still water, in $N$.\\

Next, we consider three scenarios, air cavities diffuse on top only, on bottom only, and on both top and bottom.

\smallskip

\begin{figure}[H]
\centering
\includegraphics[width=0.5\textwidth]{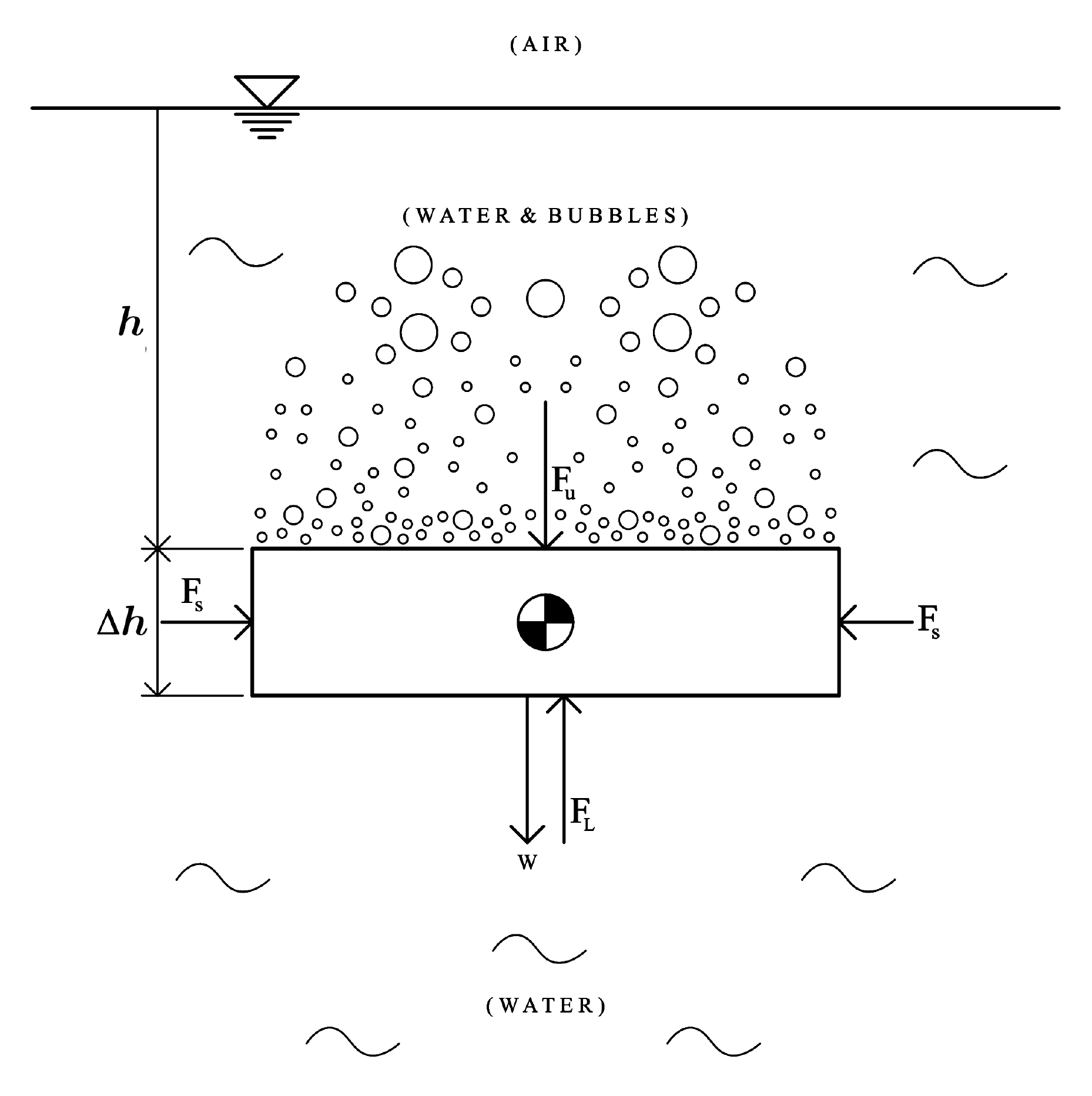}%this is figure 1.2%
\caption{\small{A cuboid submerged under water with air cavities diffusing from the top. $F_U, F_L$, and $F_S$ are the equivalent forces acting on the cuboid due to water pressure. $W$ is the weight of the cuboid.}}
\label{figure 12}
\end{figure}

{\bf Claim 1: When the body experiences air cavities diffused from the top of the cuboid as shown in Fig. \ref{figure 12}, 
then the top buoyant force will increase, causing the body to rise.}

{\it Proof 1:}\\

With cavities above the cuboid, the fluid density above it decreases; this new density of the mixture of water and bubbles is assigned a new variable $\rho_B$, and it is assumed to be uniform for future derivations. Hence, the new expression for the pressure experienced by the top side of the cuboid is shown below in \eqref{s10}. 
 \begin{equation}\label{s10}
p_U=p_{atm}+\rho_{B} g h
 \end{equation}	
 Since the bubbles originate from the top of the cuboid, the volume of space that the cuboid displaces is still that of water; hence, the pressure on the bottom side of the cuboid remains the same as that shown in \eqref{s7}.  Deriving the expression for the resulting buoyant force yields			
 
 \begin{equation}\label{s11}
F_{B_{top}}=%F_L-F_U=
p_LA-p_U A= (p_{atm}+\rho_{W} g h +\rho_W g \Delta h)A-(p_{atm}+\rho_B gh)A=(\rho_W-\rho_{B}) g h A+F_{B_{init}}
 \end{equation}
where $F_{B_{top}}$ is the buoyant force acting on top of the cuboid in still water,  in $N$.\\
 									
Hence, because $\rho_W>\rho_B$ then $F_{B_{top}}>F_{B_{init}}$.  The above derivation shows that with the bubbles diffused from the top of the cuboid, the buoyant force it experiences should increase. This upsets the vertical equilibrium and causes the body to rise. 

\smallskip

\begin{figure}[H]
\centering
\includegraphics[width=0.5\textwidth]{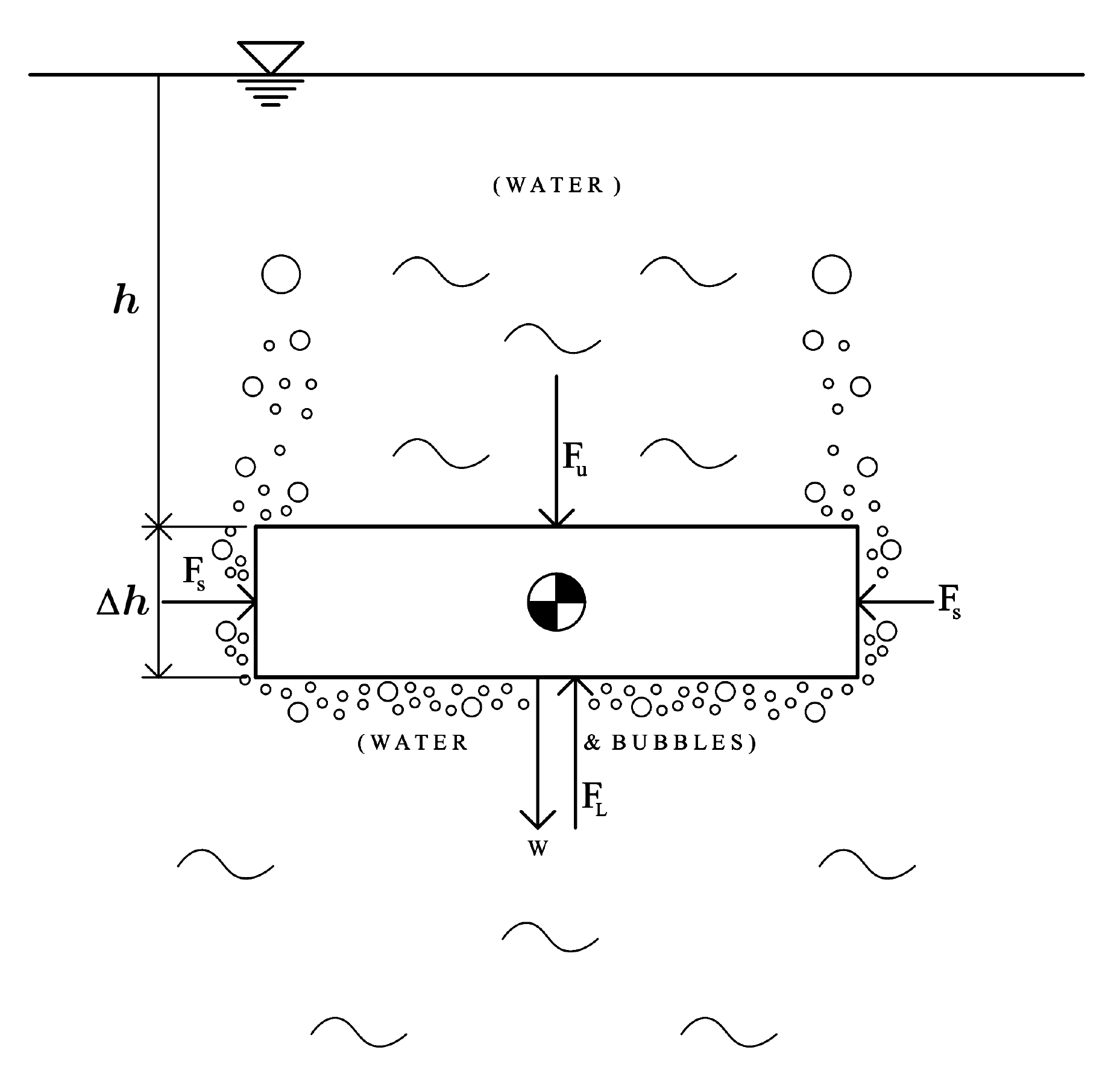}%this is figure 1.3%
\caption{\small{A cuboid submerged under water with air cavities diffusing from the bottom. $F_U, F_L$, and $F_S$ are the equivalent forces acting on the cuboid due to water pressure. $W$ is the weight of the cuboid.}}
\label{figure 13}
\end{figure}

{\bf Claim 2: When the body experiences air cavities diffused from the bottom of the cuboid as shown in Fig. \ref{figure 13}, then the bottom buoyant force will decrease, causing the body to sink.}

{\it Proof 2:}\\

In the absence of the cuboid, diffusing the bubbles in this manner will lower the density of the fluid above it. Hence, with the cuboid in its position, it is theoretically displacing the bubble-filled water. Thus, the fluid pressure underneath the cuboid can be expressed in the manner shown in \eqref{s12}. The fluid pressure above the cuboid should remain as that shown in \eqref{s6}. The derivation to find the new buoyant force follows. 
     \begin{equation}\label{s12}
p_L=p_{atm}+\rho_{W} g h+\rho_{B} g \Delta h
 \end{equation}		
 
  \begin{equation}\label{s13}
F_{B_{bttm}}=F_L-F_U=p_LA-p_U A=(p_{atm}+\rho_Wgh+\rho_{B}g \Delta h) A-(p_{atm}+\rho_Wgh)A=\rho_Bg \Delta hA
 \end{equation}
where $F_{B_{bttm}}$ is the buoyant force acting on bottom of the cuboid,  in $N$. 

Hence, because $\rho_W>\rho_B$ then $F_{B_{bttm}}<F_{B_{init}}<F_{B_{top}}$. The above derivation shows that the cuboid should sink due to lost in buoyancy when air cavities are diffused from the bottom.\\

\smallskip

\begin{figure}[H]
\centering
\includegraphics[width=0.5\textwidth]{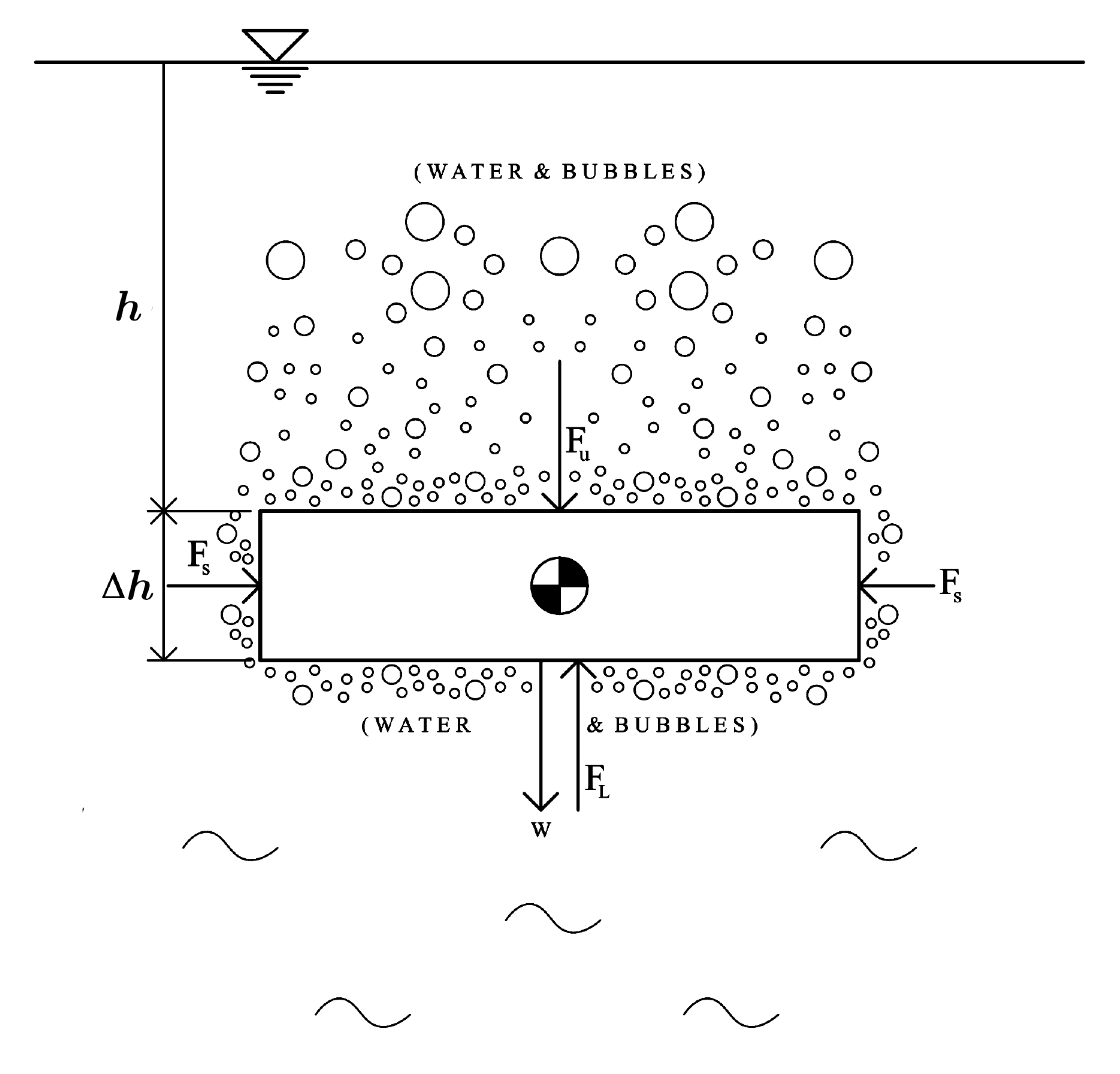}%this is figure 1.4%
\caption{\small{A cuboid submerged under water with air cavities diffusing from both top and bottom. $F_U, F_L$, and $F_S$ are the equivalent forces acting on the cuboid due to water pressure. $W$ is the weight of the cuboid.}}
\label{figure 14}
\end{figure}

{\bf Claim 3: When the body experiences air cavities diffused from both top and bottom of the cuboid as shown in Fig. \ref{figure 14} then the bottom buoyant force will decrease, causing the body to sink, similar to Claim 2.}

{\it Proof 3:}\\

In this case, it would appear as though the cuboid is submerged in a mixture of bubbles and water. Thus the expressions for the pressure at the top and bottom of the cuboid should assume the forms shown in \eqref{s10} and \eqref{s15} respectively. 
 \begin{equation}\label{s15}
p_L=p_{atm}+\rho_{B} g h+\rho_{B} g \Delta h
 \end{equation}
 
 The derivation that follows shows that the buoyant force is identical to that when bubbles are diffused from the bottom only.
  \begin{equation}\label{s16}
F_{B_{both}}=F_L-F_U=p_LA-p_U A=(p_{atm}+\rho_Bgh+\rho_{B}g \Delta h) A-(p_{atm}+\rho_B gh)A=\rho_Bg \Delta hA=F_{B_{bttm}}
 \end{equation}
where $F_{B_{both}}$ is the buoyant force acting on both top and bottom of the cuboid, in $N$. \\
 
\smallskip
 $\therefore$ we proved that 
\begin{equation}
F_{B_{both}}=F_{B_{bttm}}<F_{B_{init}}<F_{B_{top}}.
 \end{equation}

The above analysis justifies the first aspect of the hypothesis. The second aspect comes from the prediction that diffusing more air cavities results in a greater change in the fluid density, which results  in more pronounced effects on the buoyant force.

\section{Mathematical model of the vacuous cavitation bubble -- Theory}
In this section we will present the mathematical model of the vacuous bubble in water. First, we will find the radius and time of the evolution of the bubbles as parametric closed form solutions on terms of hypergeometric functions. By integration of Rayleigh equation, we will also find the  bubbles\lq{} collapsing time.

From a mathematical point of view, we may consider (i) the collapse of a spherical bubble in water, and (ii) the expansion of spherical cavity due to the pressure of the gas within the bubble.  The latter is applicable, for example, to a submarine mine.

The processes in cavitation are governed (assuming the bubble is empty and spherical) by the Rayleigh equation, see \cite{Lord,Prosp}
\be 
\rho_w\left\{R\ddot{R}+\df{3}{2}\dot{R}^2\right\}=p-P_\infty-\df{2\sigma}{R}-\df{4\mu_w}{R}\dot{R},\qquad \dot{(\,\,\,)}\equiv\df{d}{dt}\label{m0}
\en 
In (\ref{m0}), $\rho_w$ is the density of the liquid, $R=R(t)$ is the radius of a bubble, $p$ and $P_\infty$ are respectively the pressures inside the bubble and at large distance, $\sigma$ is the surface tension, $\mu_w$ the dynamic viscosity of the liquid (which in our experiment is water), and $t$ represents time.

In what follows, we shall consider an idealized case whereby the viscosity of the water is neglected, since $\mu_w\ll 1$, and we further assume that the effect of capillarity does not play an important role, which  justifies neglecting the surface tension.

\subsection{Evolution and collapse of a spherical bubble}
\begin{figure}[H]
\centering
\includegraphics[width=0.3\textwidth]{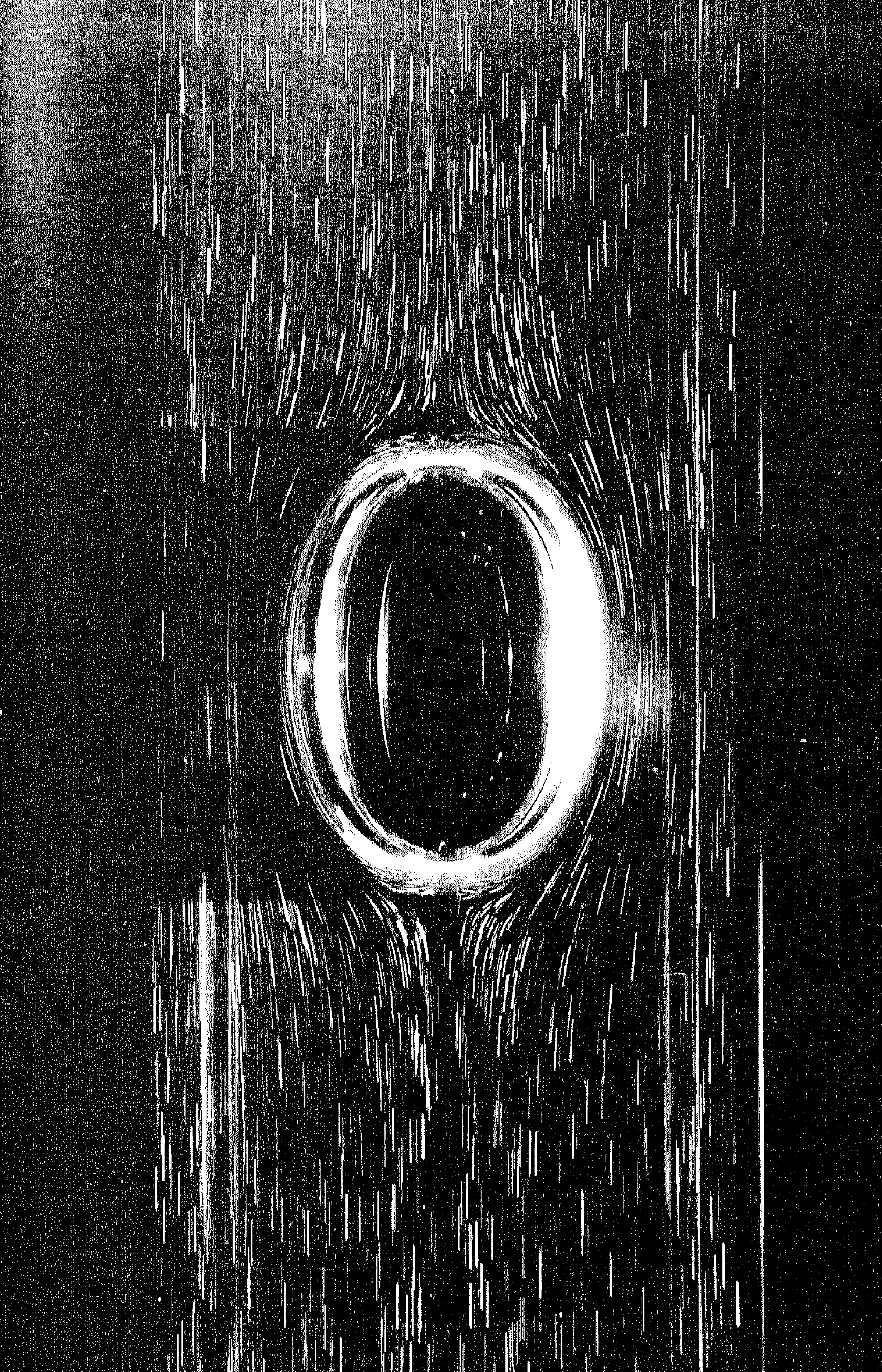}
\caption{\small{Bubble rising in water. Bubble volume is $9.7 \times 10^{-6}~ m^3$. Taken from \cite{new}}}
\label{figure15}
\end{figure}

For this case, let us consider a vacuous bubble of radius $R$ which is surrounded by an infinite uniform incompressible fluid, such as water, that is at rest at infinity, see Fig. \ref{figure15}. We remark that `infinity' in the present context refers to distance far enough away from the initial position of the bubble.

We shall assume the pressure at infinity is constant and is denoted by $P_\infty$ and for the experiment we will choose to be the atmospheric pressure.  Neglecting the body forces acting on the bubble, we will first show that
\be 
\dot{R}^2=\df{2}{3}\df{P_\infty}{\rho_w}\left[\left(\df{R_0}{R}\right)^3-1\right]\label{m1}
\en 
Consequently, we will show the time for the complete collapse of the bubble is given by
\be 
t_c=R_0\df{\Gamma(5/6)}{\Gamma(4/3)}\sqrt{\df{\pi\rho_w}{6P_\infty}}\label{m2}
\en 

To prove (\ref{m1}) and (\ref{m2}), we shall assume the motion is radial everywhere and the flow is irrotational with the velocity potential $\phi=1/r$ which satisfies the Laplace equation $\nabla^2\phi=0$.  In the expression for the velocity potential $r$ is taken to be the distance from the center of the bubble to an arbitrary point in the fluid.

Since the flow is radial, then if $\B{u}$ represent the velocity vector of the bubble, the condition of radial symmetry yields
\be 
\B{u}=(u_r, 0, 0)=-\left(\PDD{\phi}{r}, 0, 0\right)\label{m3}
\en 
Hence, 
$$u_r=-\PDD{\phi}{r}=\df{1}{r^2}.$$

Thus, on the surface of the bubble
\be 
u_r=\df{m}{R^2}=\dot{R}\label{m4}
\en 
and therefore the velocity potential becomes
\be 
\phi=\df{\dot{R}R^2}{r}\label{m5}
\en 

Using Bernoulli\rq{}s equation,
\be 
\df{p}{\rho}+\df{1}{2}|\B{\nabla}u|^2-\PDD{\phi}{t}=C(t)\label{m5a}
\en 
then upon substituting for the respective terms in (\ref{m5a}), we obtain
\be 
\df{p}{\rho_w}+\df{1}{2}u_r^2-\df{1}{r}\left(2R\dot{R}^2+R^2\ddot{R}\right)={\cal A}(t)\label{m6a}
\en 
Next, using the condition at infinity, namely $r=\infty$, we have that
$\df{P_\infty}{\rho_w}=const.$

Thus, equation \eqref{m0} becomes
\be 
\df{p}{\rho_w}-\df{3}{2}\dot{R}^2-R\ddot{R}=\df{P_\infty}{\rho_w}\label{m7}
\en
But, as we have assumed the bubble is vacuous, then $p=0$ inside the bubble and therefore equation (\ref{m7}) reduces further to
\be 
3\dot{R}^2+2R\ddot{R}=-2\df{P_\infty}{\rho_w}\label{m8}
\en 

Multiplying (\ref{m8}) through by $R^2\dot{R}$ and integrating, we get
$$
R^3\dot{R}^2=-\df{2}{3}\df{P_\infty}{\rho_w}R^3+{\cal B}.
$$
Since we use initial conditions $R(0)=R_0$ and  $\dot{R}(0)=0$ we find the integration constant
$$
{\cal B}=\df{2}{3}\df{P_\infty}{\rho_w}R_0^3.
$$
Therefore, we see at once that 
\be
\dot{R}^2=\df{2}{3}\df{P_\infty}{\rho_w}\left[\left(\df{R_0}{R}\right)^3-1\right]\label{imp}
\en
which establishes the result given by equation (\ref{m1}).

This equation is very interesting since up to a factor  it represents a conservation law for the dynamics of the radius of the bubble, and it deserves its own attention. To proceed, we will use the set of transformations, given by Kudryashov, see \cite{Kud}, namely
$R=S^\epsilon, dt=R^\delta d \tau$, where $\epsilon, \delta$ are constants that depend on the dimension of the bobble, and $S, \tau$ are the new dependent and independent variables.

Using the transformations on \eqref{imp}, we obtain the new dynamics in $S$ and $\tau$
\be
S_\tau^2=\frac{2}{3}\frac{P_\infty}{\rho_w}\frac{1}{ \eps^2}(R_0^3~S^{-3 \eps}-1)S^{-2(\eps-1-\eps\delta)}\label{imp2}
\en

To find $S$ one sets $\eps=\frac{1}{N}$, and $\delta=N+1$, where $N=3$ is the dimension of the bubble, which will in turn reduce \eqref{imp2} to the simple equation
\be
S_\tau=3 S \sqrt {\frac{2}{3}\frac{P_\infty}{\rho_w}}\sqrt{R_0^3~S-S^2}\label{imp3}
\en
By integrating the above with $S(0)=R_0^3$ we obtain the rational solution
\be
S(\tau)=\frac{R_0^3}{M\tau^2+1},
\en
where for convenience we let $M=\frac 3 2 \frac{P_\infty}{\rho_w}R_0^6.$
Once we have $S$, we can  find the parametric solutions for the bubble radius $R(\tau)$ and evolution time of the bubble $t(\tau)$ and we get

\begin{eqnarray}\label{ze}
R(\tau)&= \frac{R_0}{(M\tau^2+1)^{\frac 1 3}}~,\\ \notag
t(\tau)&=R_0^4\int_0^{\tau} \frac{d \xi}{(M \xi^2+1)^{\frac 43}}
\end{eqnarray}
The integral for the evolution of the time for bubble  can be calculated analytically in terms of hypergeometric functions to give
\be
t(\tau)=\frac{\tau}{2}\left[\frac{3}{(M \tau^2+1)^\frac 1 3}-{}_2 F_1\left(\frac 1 3, \frac 1 2 ; \frac 3 2; -M \tau^2\right)\right]
\en

\bigskip
Next, we will find the time for the total collapse $t_c$ of the bubble by  integration of  equation (\ref{m1}), which yields to
\be 
t_c=R_0^{-3/2}\sqrt{\df{3}{2}\df{\rho_w}{P_\infty}}\int_0^{R_0}\df{R^{3/2}}{\sqrt{1-\big(\frac{R}{R_0}\big)^3}}\,dR\label{m9}
\en 
If we let $R=R_0\sin^{2/3}\theta$, where $\theta\in[0, \pi/2]$ then the integral (\ref{m9}) transforms to 
$$
t_c=\df{2R_0}{3}\sqrt{\df{3}{2}\df{\rho_w}{P_\infty}}\int_0^{\pi/2}\sin^{2/3}\theta\,d\theta
$$
which by comparing with the integral relation for Beta function, namely,
$$
{\rm B}(m,n)=2\int_0^{\pi/2}\cos^{2m-1}\theta\sin^{2n-1}\theta\,d\theta
$$

we obtain
\be 
t=\df{R_0}{3}\sqrt{\df{3}{2}\df{\rho_w}{P_\infty}}{\rm B}(1/2,5/6)=\df{R_0}{3}\sqrt{\df{3}{2}\df{\rho_w}{P_\infty}}\df{\Gamma(1/2)\Gamma(5/6)}{\Gamma(4/3)}\label{m10}
\en 
Finally, using the relationship between Beta and Gamma functions, and noting that  
$\Gamma(1/2)=\sqrt{\pi}$, we obtain from (\ref{m10}) the total time of collapse of the bubble to be that given by (\ref{m2}).

It is evident, the total collapse time is invers  proportional to square root of the pressure at infinity.  Now, as an example if we take, in c.g.s. units, $\rho_w=1$ g/cm$^3$, $R_0=1$ cm and $P=1$  atmosphere, then from (\ref{m2}) we see that the total time of collapse of such a bubble is $t=0.000915$ sec.  Moreover, using \eqref{imp} the kinetic energy at any instant is given by
$$
2\pi \rho_w R^3\dot{R}^2=\tf{4}{3}\pi P_\infty(R_0^3-R^3)
$$
Thus, when the collapse occurs $R(t_c)=0$, the energy-transfer to water is $E_c=\tf{4}{3}\pi P_\infty R_0^3$.

\subsection{Expansion of spherical cavity}

As in the case (i), equations (\ref{m5}) and (\ref{m5a}) are still applicable for the problem of the expanding cavity.  However, in this case, we neglect the pressure $P_\infty$ at infinity.

Taking $\mathcal{P}$ to be the initial pressure on the cavity, when $R=R_0$, and $\dot{R}=0$ then, assuming the adiabatic law of expansion, the internal pressure at time $t$ is given by
$$
\df{p}{\mathcal{P}}=\left(\df{R_0}{R}\right)^{3\gamma},
$$
where $\gamma$ is the constant adiabatic index.

Hence, we now have 
\be 
3\dot{R}^2+2R\ddot{R}=2\mathcal{C}^2\left(\df{R_0}{R}\right)^{3\gamma}\label{m12}
\en 
where $\mathcal{C}=\sqrt{\mathcal{P}/\rho_w}$  has the dimension of a velocity and determines how rapidly the bubble changes its shape.

Equations \eqref{imp}, \eqref{m12} are very important since they are related to the Einstein–Friedmann
dynamical equations of barotropic FRW cosmologies with a cosmological constant where the adiabatic index takes the values of $\gamma=0,1,4/3$ for a universe dominated by vacuum, pressure-less mater, and radiation respectively, see \cite{Man1,Man2}.

Using the method outlined in the case (i), the integration of (\ref{m12}) yields
\be 
\dot{R}^2=\df{2 \mathcal C^2}{3(\gamma-1)}\left[\left(\df{R_0}{R}\right)^3-
\left(\df{R_0}{R}\right)^{3\gamma}\right]\label{m13}
\en 
%This equation determines the initial acceleration, $\ddot{R}$, in the radius is $\mathcal{C}^2/a^2$, regardless the law of expansion used.  
It is easy to see that from (\ref{m12}) and (\ref{m13}) $\dot{R}$ has the maximum when
$$
\left(\df{R}{R_0}\right)^{3(\gamma-1)}=\gamma
$$
and the minimum when
$$
\left(\df{\dot{R}}{\mathcal{C}}\right)^2=\df{2}{3\gamma^{\gamma(\gamma-1)}}
$$

We remark that the solution is not easily obtained for an arbitrary value of $\gamma$, for which, in general, one may obtain solutions in terms of Weierstrass elliptic functions, see \cite{Kud}, or hypergeometric functions, see \cite{Man2}. 
Notice that in the case of $\gamma=0$, \eqref{m13} is simply \eqref{imp} with solution given by \eqref{ze} and $M=-\frac 3 2 \mathcal C^2 R_0^6$, while for $\gamma=1$ implies that $\dot R=0$ which shows that the bubble stays constant. In the later case $\gamma=4/3$ from \eqref{m13} we obtain

\be 
\dot{R}^2=2 \mathcal C^2 \left(\df{R_0}{R}\right)^3 \left[1-
\df{R_0}{R}\right]\label{m14}
\en

Since $R_0 \le R(t)$ let  $\frac {R(t)}{R_0}=1+\chi(t)$ in \eqref{m14} to obtain
\be
(1+\chi)^2\dot \chi=\df{\mathcal{C}}{R_0}\sqrt{2\chi} \label{m14a}
\en
Thus, upon integration and using $\chi(0)=0$ we get
$$
\sqrt \chi \left(1+\df{2}{3}\chi+\df{1}{5}\chi^2\right)=\frac{\sqrt 2\mathcal C}{2R_0}t
$$
Restoring to original variables we find
$$
t=\df{\sqrt2}{15\mathcal{C}R_0}\sqrt{\df{R}{R_0}-1}\left(8R_0^2+4RR_0+3R^2\right)
$$

Now, if the initial diameter of the cavity is 100 cm, and the initial pressure $\mathcal{P}=1000$ atmosphere, then $\mathcal{C}=3.16\times 10^4$ cm/sec.  Thus, the radius of the cavity is double in $\tf{1}{260}$ of a seconds, and multiplied five-fold in about $\tf{1}{30}$ sec.  Hence, the initial acceleration of the radius is $2\times 10^7$ cm/sec$^2$, and this justifies the neglect of gravity in the initial stages of the motion.

Note that, the maximum of $\dot{R}$ occurs when $R/R_0=4/3, t=0.0016$ sec, 
and hence is approximately 14500 cm/sec,
which is about $\tf{1}{10}$ of speed of sound in water.  Thus, we conclude for initial pressure of the order of 10000 atmospheres (or more), the obtained speed is comparable with the speed of sound.  In such cases the effect of compressibility should not be neglected and the above theory will not yield an accurate results.

\section{Experiments}
\subsection{Setup}
\begin{figure}[H]
\centering
\includegraphics[width=0.5\textwidth]{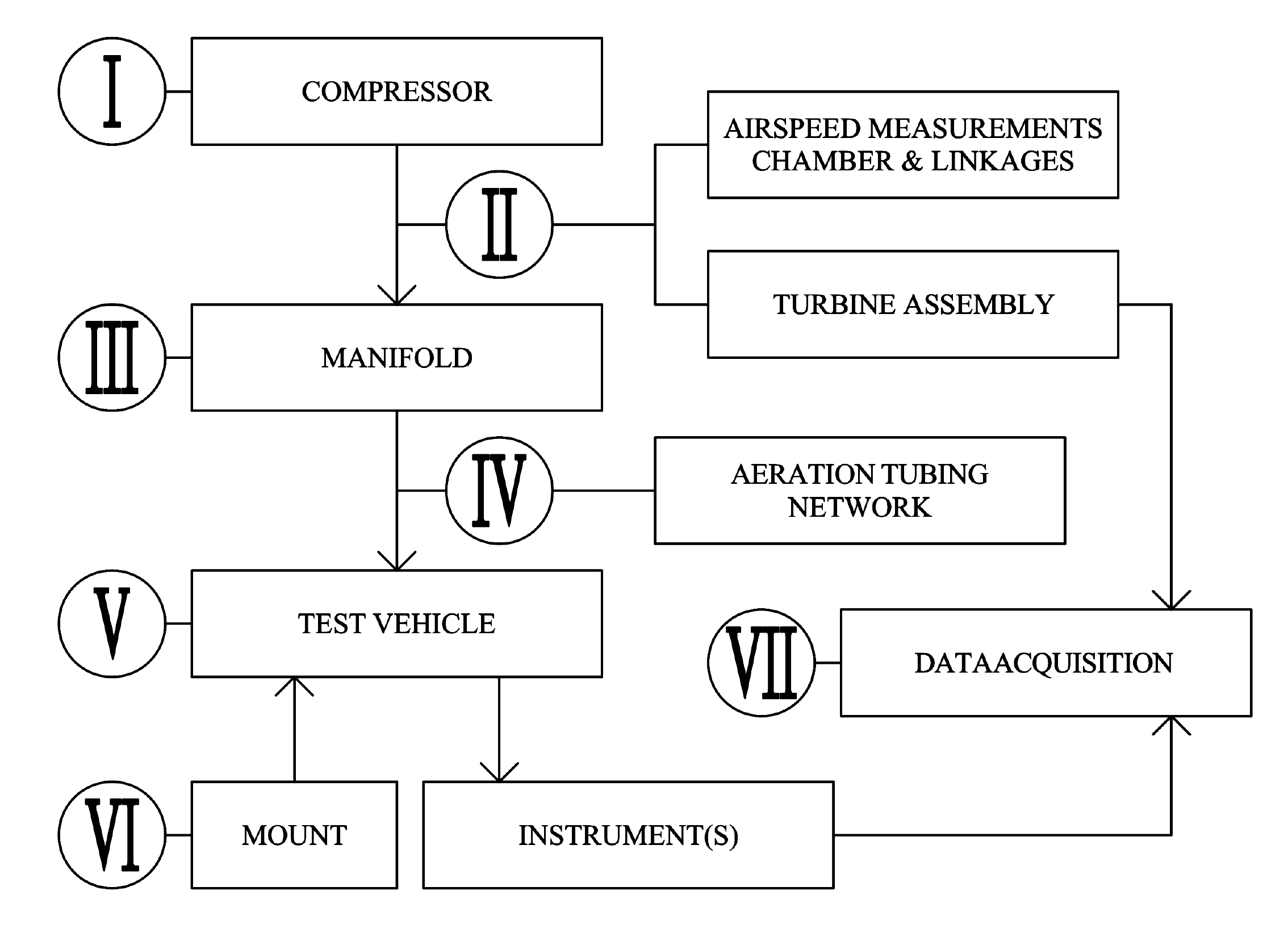}
\caption{\small{A schematic showing the flow of compressed air and experimental data.}}
\label{figure 1}
\end{figure}

\begin{figure}[H]
\centering
\includegraphics[width=0.5\textwidth]{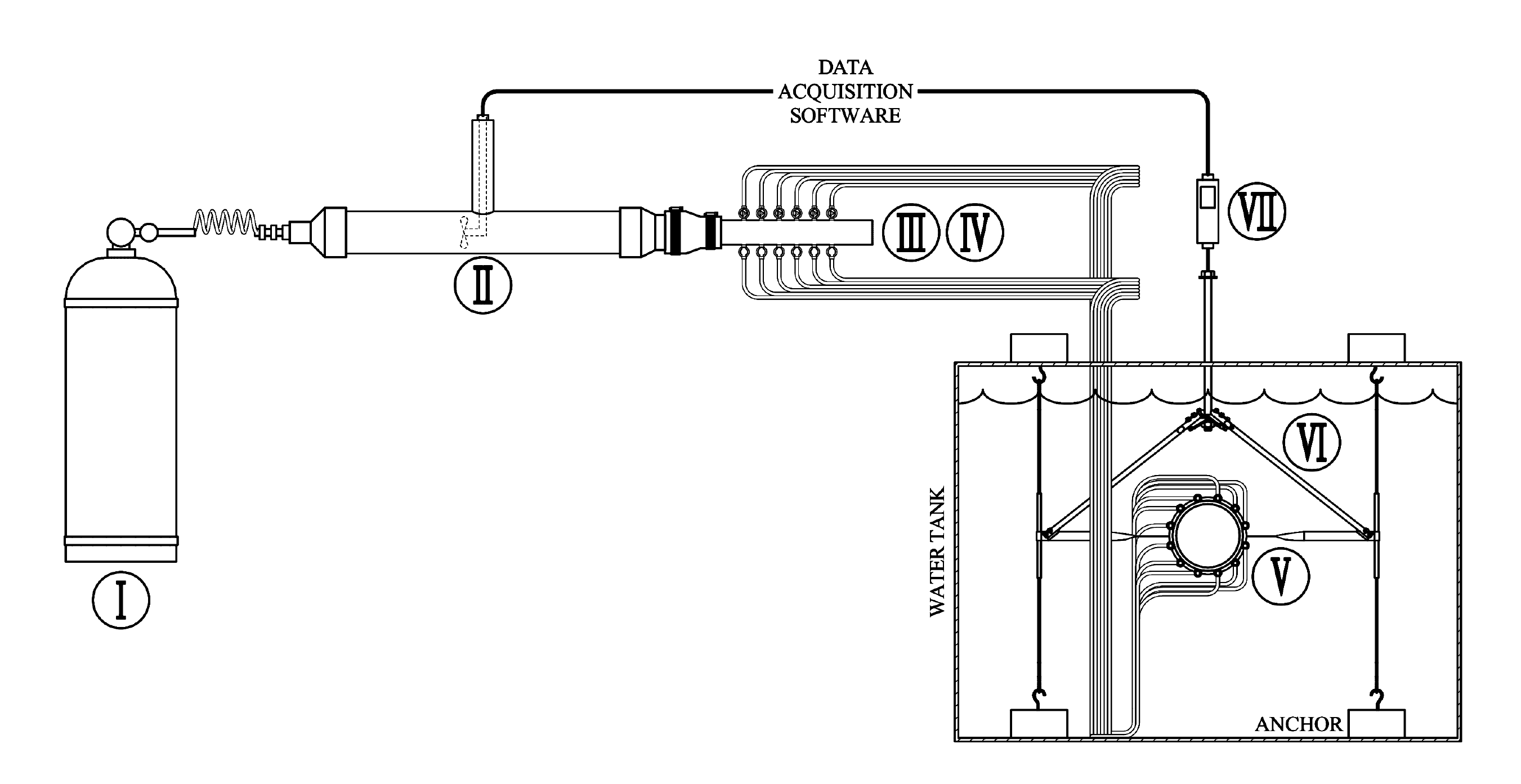}
\caption{\small{A sketch of experimental setup.}}
\label{figure 2}
\end{figure}

\begin{figure}[H]
\centering
\includegraphics[width=0.5\textwidth]{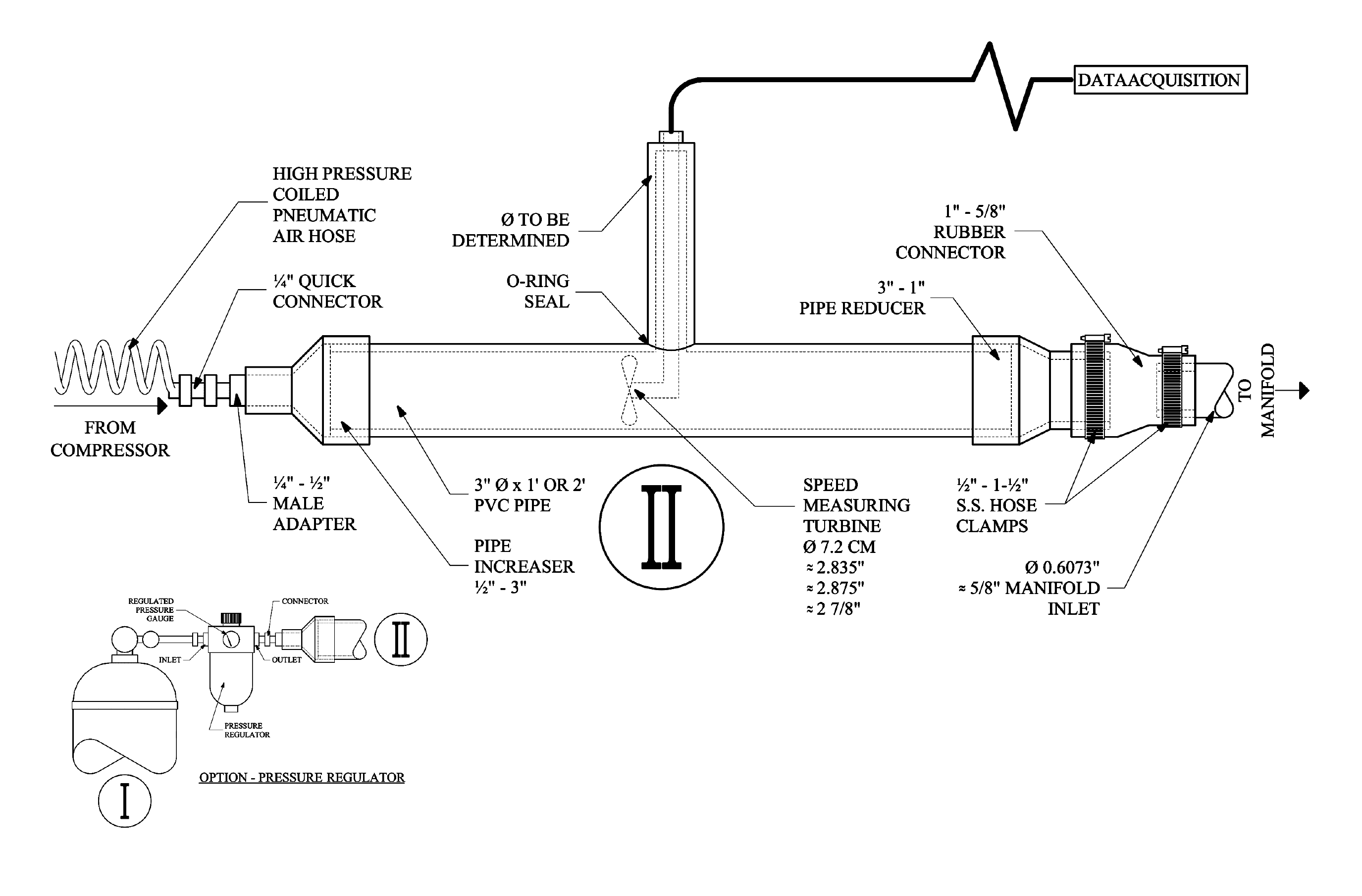}
\caption{\small{A sketch of the velocity measurement chamber.}}
\label{figure 3}
\end{figure}

\begin{figure}[H]
\centering
\includegraphics[width=0.5\textwidth]{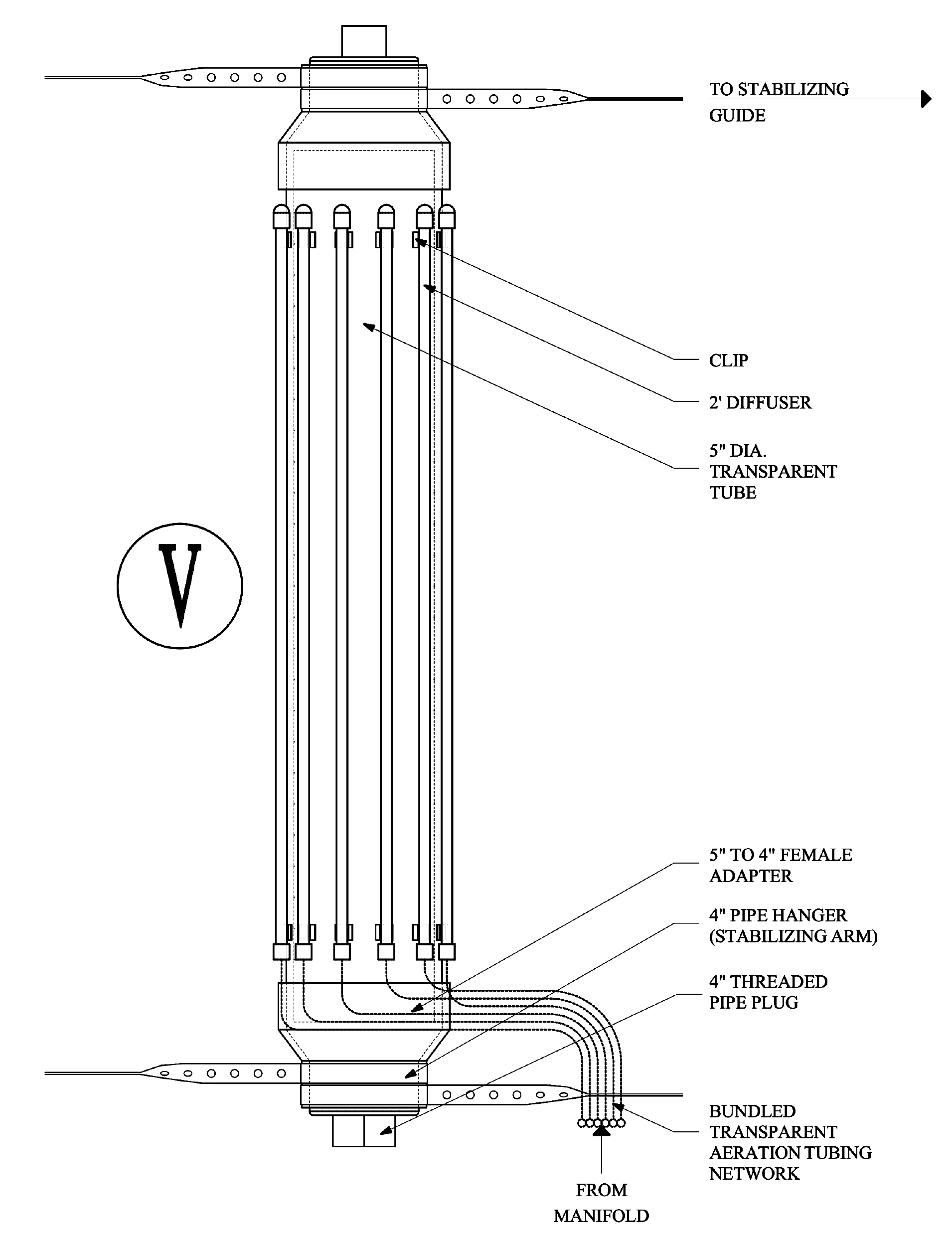}
\caption{\small{A sketch of the top view of the test vehicle. The number of diffusers shown here is for illustrative purposes only. Sixteen diffusers were used in reality, divided equally between the top and the bottom of the vessel.}}
\label{figure 4}
\end{figure}

\begin{figure}[H]
\centering
\includegraphics[width=0.5\textwidth]{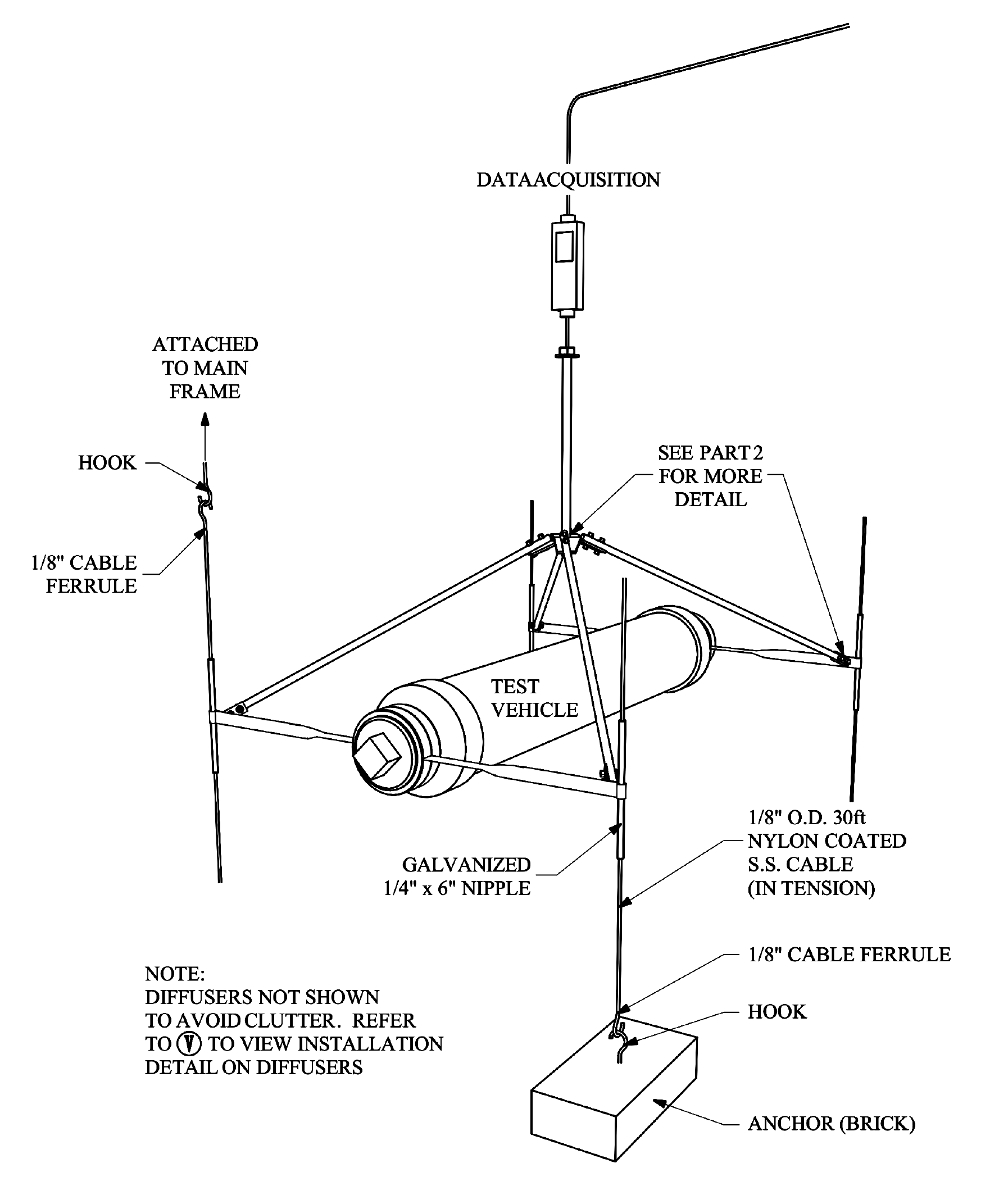}
\caption{\small{A sketch of the test vehicle suspended by a force gauge.}}
\label{figure 5}
\end{figure}

This research would indicate if changing surrounding fluid density could indeed affect an underwater vehicle's rate of diving and surfacing. Following the schematic shown in Figs. \ref{figure 1}- \ref{figure 5}, a heavy-duty industrial air compressor delivered compressed air to the diffusers attached to the test vehicle via a network of aeration tubing system. 
Here we assume that the body is a cuboid in a hydrostatic fluid, and that the fluid has uniform density. For an area $A$ the pressure exerted by the water on the cuboid can be expressed as shown in equation \eqref{m1}, see \cite{Yon}.

	Fig. \ref{figure 2}  illustrates that immediately downstream of the compressor is the velocity measurement chamber shown below in Fig. \ref{figure 3}. This chamber was constructed from a PVC pipe of known inner diameter. With the air velocity measured, this enabled the calculation of the volumetric flow rate of the compressed air, which was a direct measurement of the amount of air bubbles leaving the test vehicle. In reality, instead of a turbine as shown in the figure, a hot-wire anemometer was used as the velocity measuring instrument for enhanced accuracy.

Downstream, the compressed air is routed to flexible diffusers through two eight-outlet steel manifolds. One manifold was connected to 8 diffusers attached to either the top or the bottom of the test vehicle.  The submarine was constructed such that it would sink when unsupported. 
%Supplying the bottom diffusers with compressed air would create micro cavitation bubbles underneath the vessel, theoretically making it sink faster. On the other hand, applying compressed air to the top diffusers should induce a buoyant force and make the vehicle sink slower, or depending on the pressure differential it may cause the vessel to even rise. 
To verify the correlation between the volume of air diffused into the surrounding water of an underwater vessel and the change in the buoyant force acting on it we used two instruments for measurement purposes: a hot-wire anemometer and a force gauge which measured the buoyancy forces.

Referring to Fig. \ref{figure 5}, the test vehicle was suspended underwater by a force gauge inside the water tank of the Nonlinear Waves Research Laboratory\cite{NLWL}. The force gauge was then supported by a frame bolted directly to the main structure of the water tank. The vehicle had guides equipped at each corner to limit its motion to the vertical direction (sliding up and down steel rods). The overall density of the vehicle is greater than that of the surrounding water, causing it to sink by default. 

\subsection{Procedure}

We have performed three experiments: first with only the top diffusers ejecting air bubbles, second with only the bottom, and third with all the diffusers activated (top  and bottom). The reading on the hot-wire anemometer was the velocity in $m/sec$ (the independent variable). The dependent variable was the buoyancy of the test vessel as sensed by the force gauge, and was measured  in $N$. With the experiment set up, and the test vehicle fully submerged in the tank, the force gauge was zeroed while fully supporting the vessel prior to each trial. When only the top diffusers were activated, the force gauge should register negative forces, indicating an increase in buoyancy. The reverse should be true for the case when only the bottom diffusers were activated.

The air compressor used for this experiment was capable of supplying compressed air at $120~psi$, ($1~psi=6894.8~ Pa.$) However, this pressure could not be sustained for long due to compressor limitations. With the valve fully opened for a prolonged period, test trials indicated that the pressure would drop gradually to about $40 ~psi$ in the span of about two minutes. The procedure of this experiment had been designed to accommodate this limitation. 

Initially, the compressor was allowed to run until the compressed air reached an indicated static pressure of $120~psi$ on the compressor's analog pressure gauge. The experiment began when the valve of the air compressor was fully opened at this indicated pressure. The valve remained open for two minutes while the compressor reservoir lost pressure steadily. During this time, the air velocity was measured by the hot-wire anemometer while the buoyant force was measured by the force gauge through a computer software that recorded data at one second intervals. The data was recorded on a computer, then tabulated, graphed and analyzed. 
\section{Results}
\subsection{General observations}

\begin{figure}[H]
\centering
\includegraphics[width=0.5\textwidth]{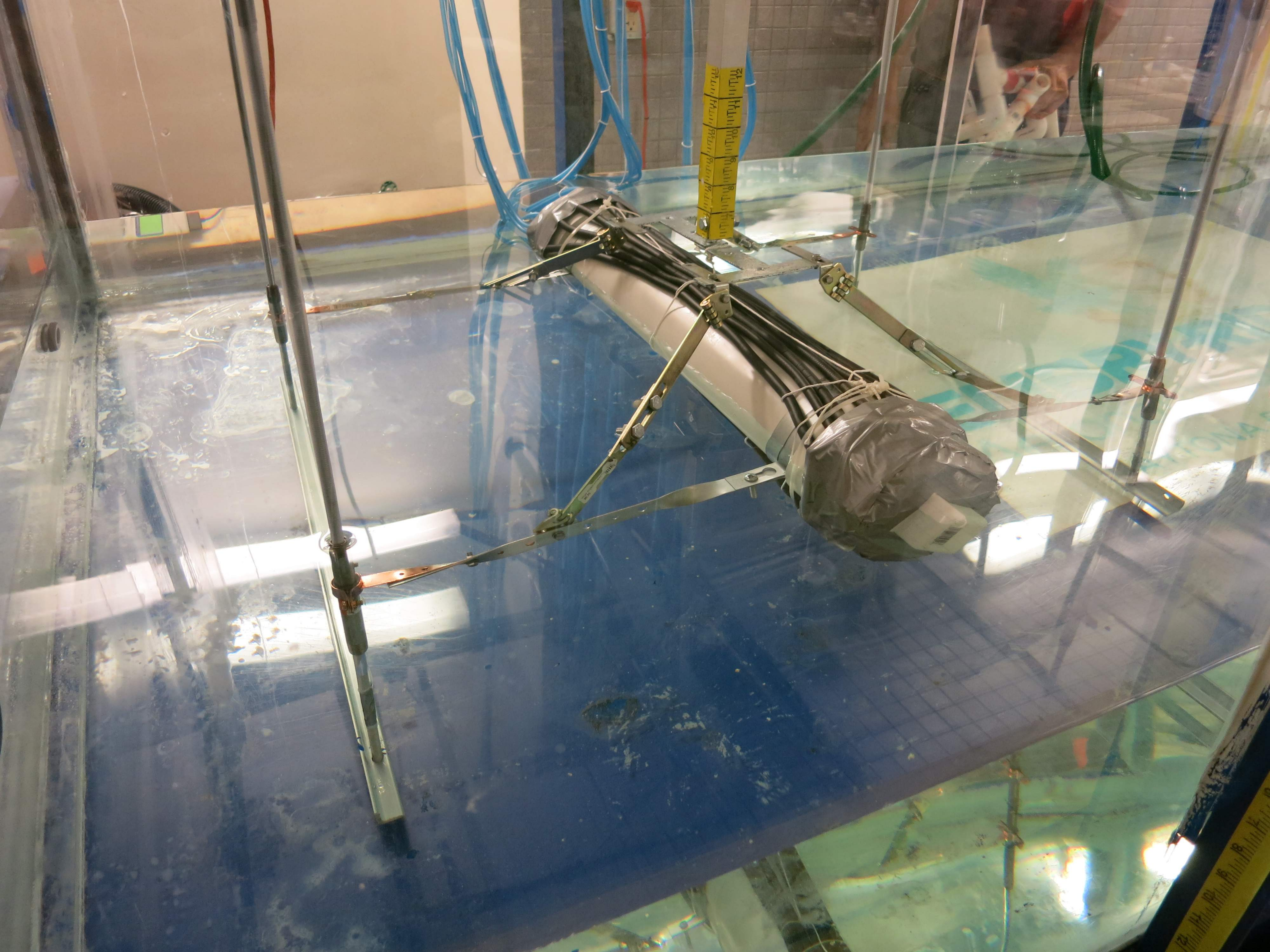}
\caption{\small{The test vehicle suspended by the force gauge while water was filling up the water tank. The black flexible diffusers visible on the top of the test vehicle were not yet activated. The vehicle was stabilized by four arms guided vertically along steel rods mounted directly to the main structure of the water tank as detailed in the sketches of the experimental setup.}}
\label{figure 6}
\end{figure}

\begin{figure}[H]
\centering
\includegraphics[width=0.5\textwidth]{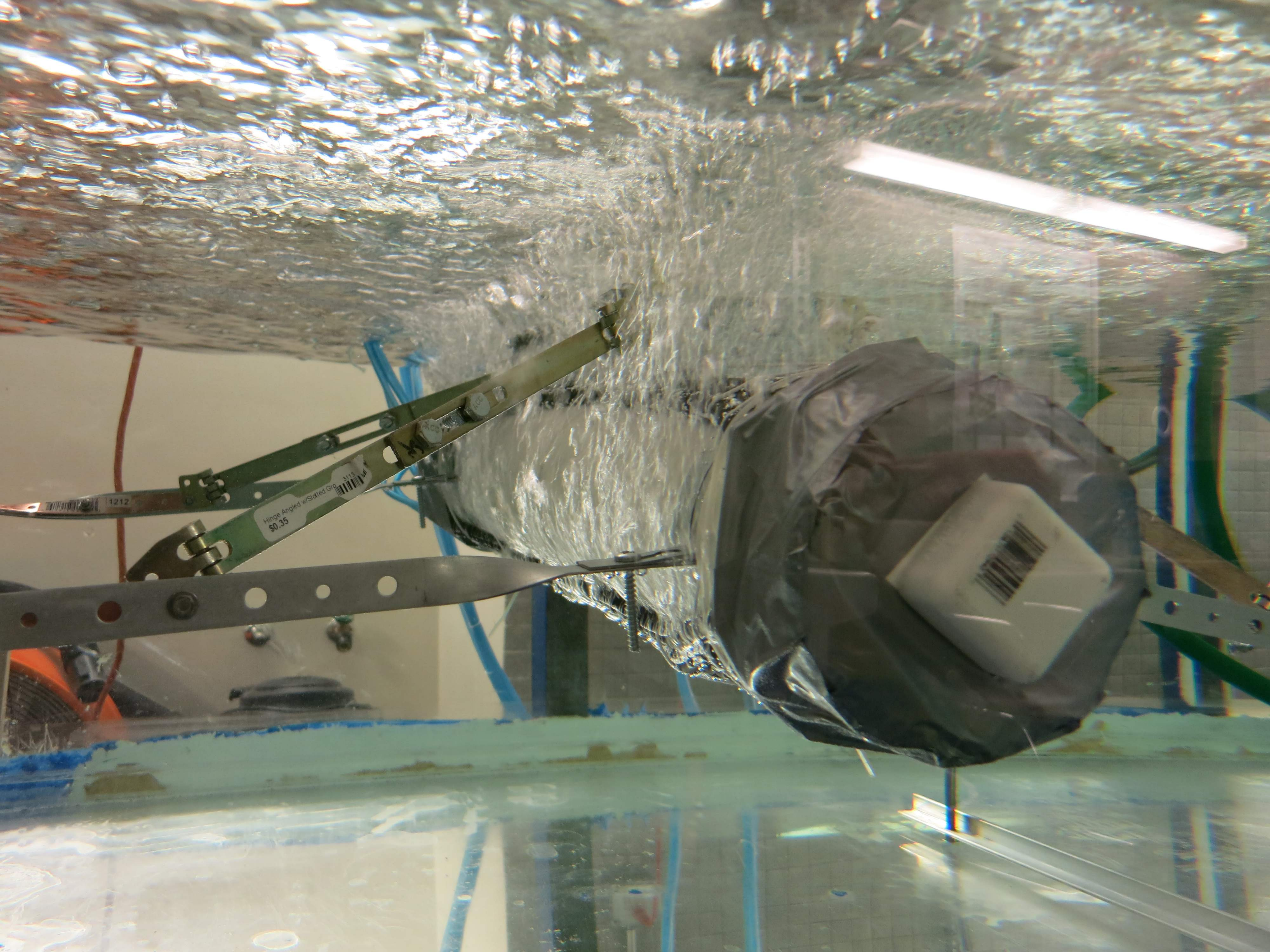}
\caption{\small{This picture depicts an ongoing trial in which both the top and bottom diffusers were activated. Air cavities were observed rising to the surface in a rapid fashion.}}
\label{figure 7}
\end{figure}

\begin{figure}[H]
\centering
\includegraphics[width=0.5\textwidth]{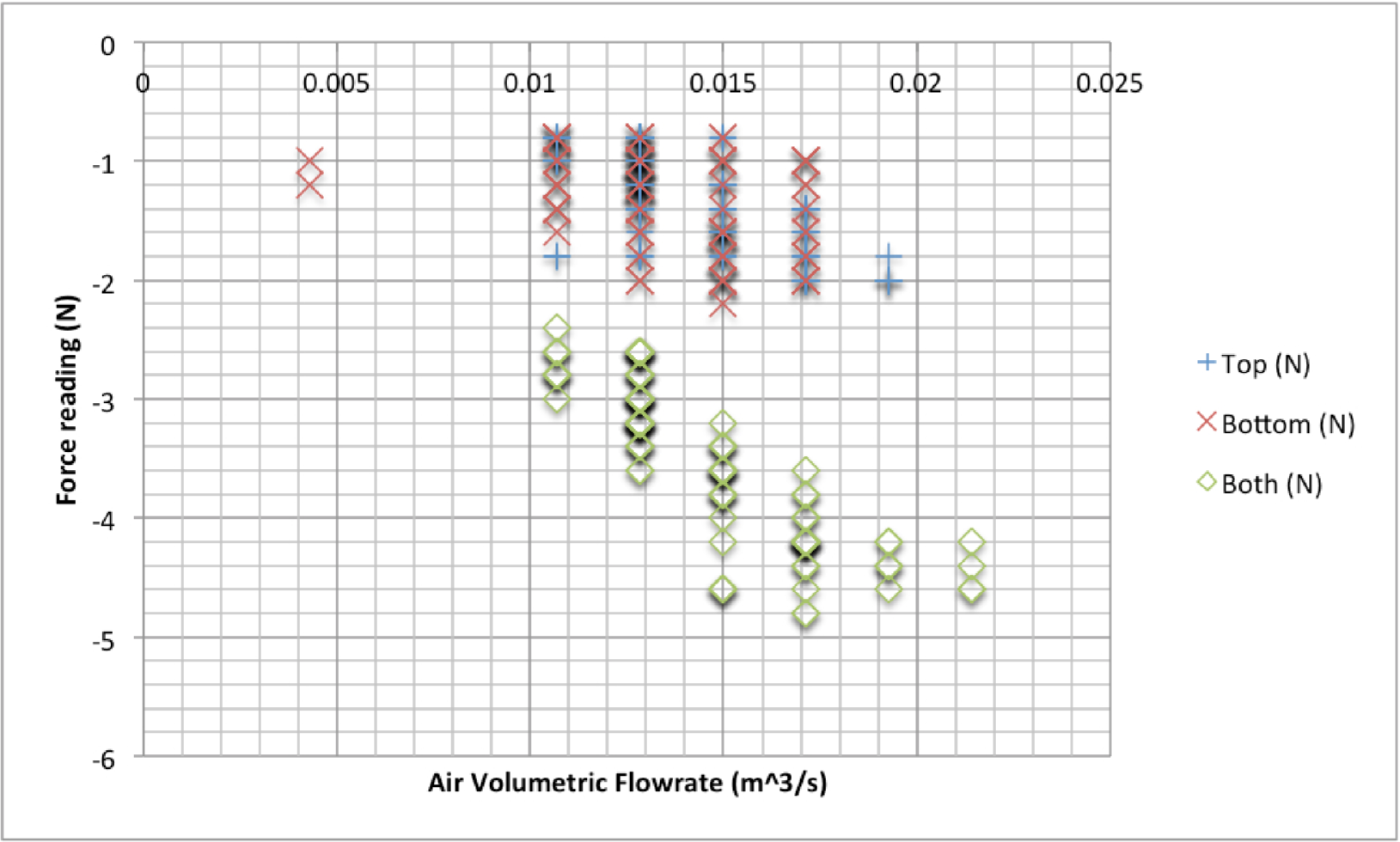}
\caption{\small{Chart showing negative forces registered by the force gauge when air bubbles were diffused into the surrounding water.}}
\label{figure 8}
\end{figure}

Contrary to the hypothesis, the underwater vessel experienced an increase in buoyancy regardless of whether the top or the bottom diffusers were activated. The following chart from Fig. \ref{figure 8} represents the buoyant force registered by the force gauge during the experiment. There were 120 scatter-points plotted for each trial (top, bottom, or both). Each scatter point represents the state of air volumetric flow rate and the resulting buoyant force at each second of a trial. The chart shows that whenever air bubbles were diffused into the water surrounding the test vehicle, a negative force was induced, which indicates an increase in buoyancy. 

Apparently, the chart also shows that at any given flow rate, the force gauge registered high amount of fluctuations in the induced buoyant force. This may be a hint that if the test-vehicle were unsupported, diffusing the air bubbles would result in violent turbulence and instability. 

\begin{figure}[H]
\centering
\includegraphics[width=0.5\textwidth]{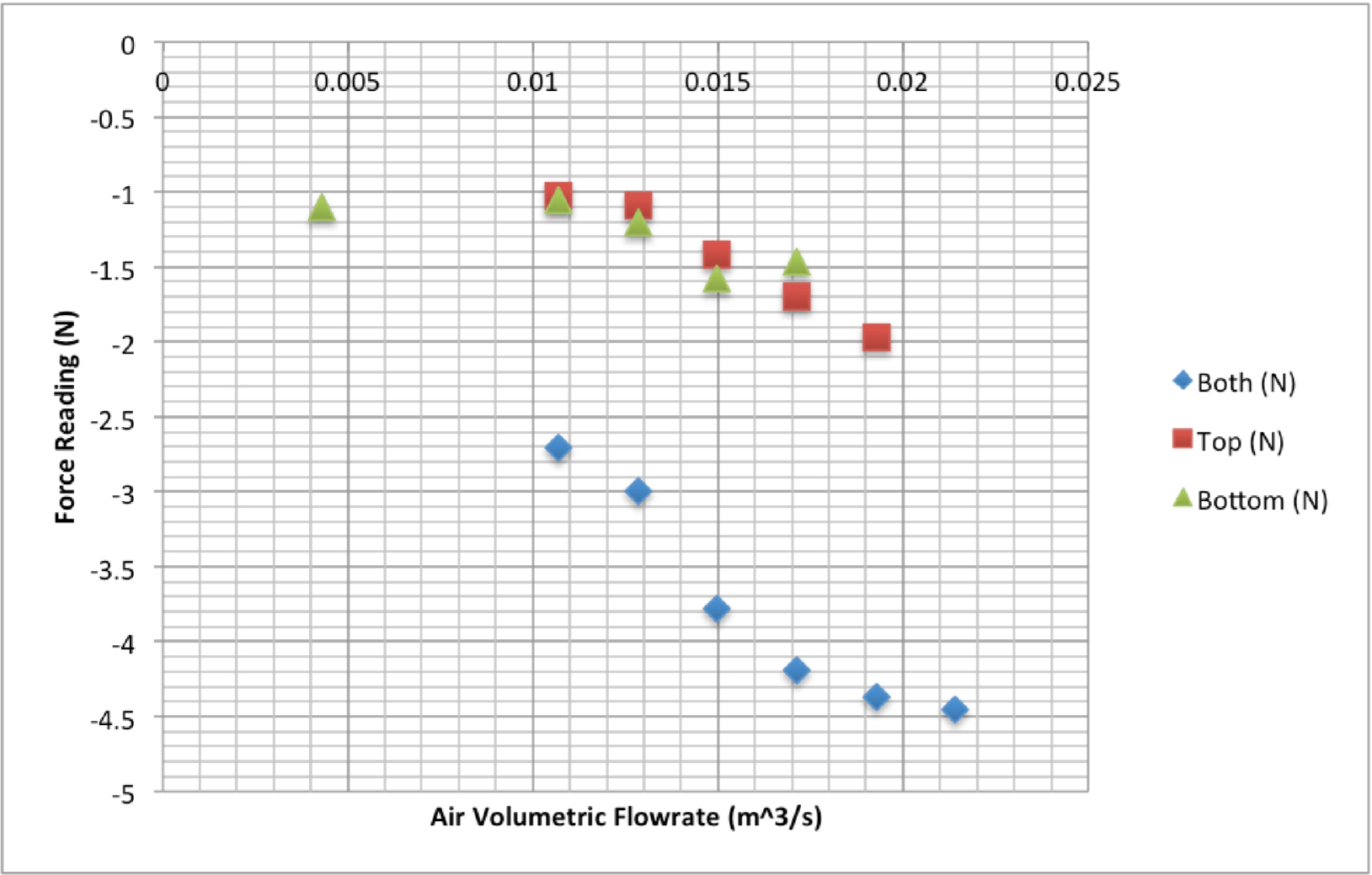}
\caption{\small{Chart representing the average buoyant forces experienced by the underwater vessel at any given air volumetric flow rate during the experiment.}}
\label{figure 9}
\end{figure}

It was difficult to establish a trend because of the high amount of spread in the scatter points due to the fluctuations in the force readings. Hence, averages of the force readings were computed for each flow rate, see Fig. \ref{figure 9}. 

\subsection{Proving or Disproving Hypothesis Part-1}
It is interesting to note that not only the induced forces were negative for both independent  cases (top on, bottom off and vice versa), but also that their magnitudes were approximately equal. The initial hypothesis was that as the density of the water column above the underwater vessel decreases due to dilution with air bubbles, the buoyant force acting on the vessel increases. Similarly, when the density of the water just underneath the vehicle is reduced by activating the bottom diffusers, the vehicle should experience less buoyant force, causing it to lose buoyancy. The evidence presented by this experiment seemed to disagree with the hypothesis. 

However, there is a plausible cause that could explain the phenomenon observed. In the case that only the top diffusers were active, the overhead water column decreased in density, therefore causing the force gauge to display negative forces, which agreed with the prediction. 

In the case of only the bottom diffusers being active, the bubbles may have decreased the density of the water directly underneath the vessel, but the same bubbles would soon rise and also dilute the overhead water column. The former event could have produced negligible effects as the latter took place. It would cause the density of the water column above the vessel to also decrease just as it would with the case of only the top diffusers being active. In effect, almost identical results were observed for both cases. Therefore, this result could not prove or disprove this aspect of the hypothesis conclusively. 

\subsection{Proving or Disproving Hypothesis Part-2}

Another deviation from the original hypothesis was that the magnitude of the force readings seemed to be independent of the amount of air bubbles released into the surrounding water per unit time. Fig.  \ref{figure 8} indicates that when only either the top or the bottom diffusers were active, there was an upward trend that the more the air bubbles diffused, the larger the induced buoyant force. That would be consistent with the hypothesis. However, the case when both the top and bottom diffusers were activated proved otherwise. For the same amount of air volumetric flow rate, the induced buoyant force seemed to double. This phenomenon clearly proved that the amount of buoyant force induced was not affected by the amount of air released to the surrounding water, but rather on the number of active diffusers. In other words, it is possible to conjecture at this point that the greater the number of diffusers activated, the greater the resulting induced buoyant force, and vice versa. 

\smallskip
To explain this pivotal observation, the following general form of fluid linear momentum equation shown below was considered. 
\begin{equation}\label{p}
\sum \vec{F}=\frac{d}{dt}\oiiint \rho \vec{  V}d \mathcal V+\oiint \rho \vec{V} ~\vec{V}\cdot d \vec{\mathcal S}
\end{equation}

\noindent  where $\sum \vec{F}$ is sum of all external forces acting on the control volume, in $N$, $\rho$ is the density of fluid leaving or entering the control volume (air cavities),  in $kg/m^3$, $\vec{V}$ =velocity of fluid entering or leaving the control volume (also air),  in $m/sec$,
$\vec{V}\cdot d \vec{\mathcal S}=\vec{V}\cdot \hat {n} ~d \mathcal S$ is the velocity component of fluid entering or leaving the control volume perpendicular to the boundary, in $m^3/sec$, and
$\mathcal V$ size of the control  volume, in $m^3$.\\

\begin{figure}[H]
\centering
\includegraphics[width=0.5\textwidth]{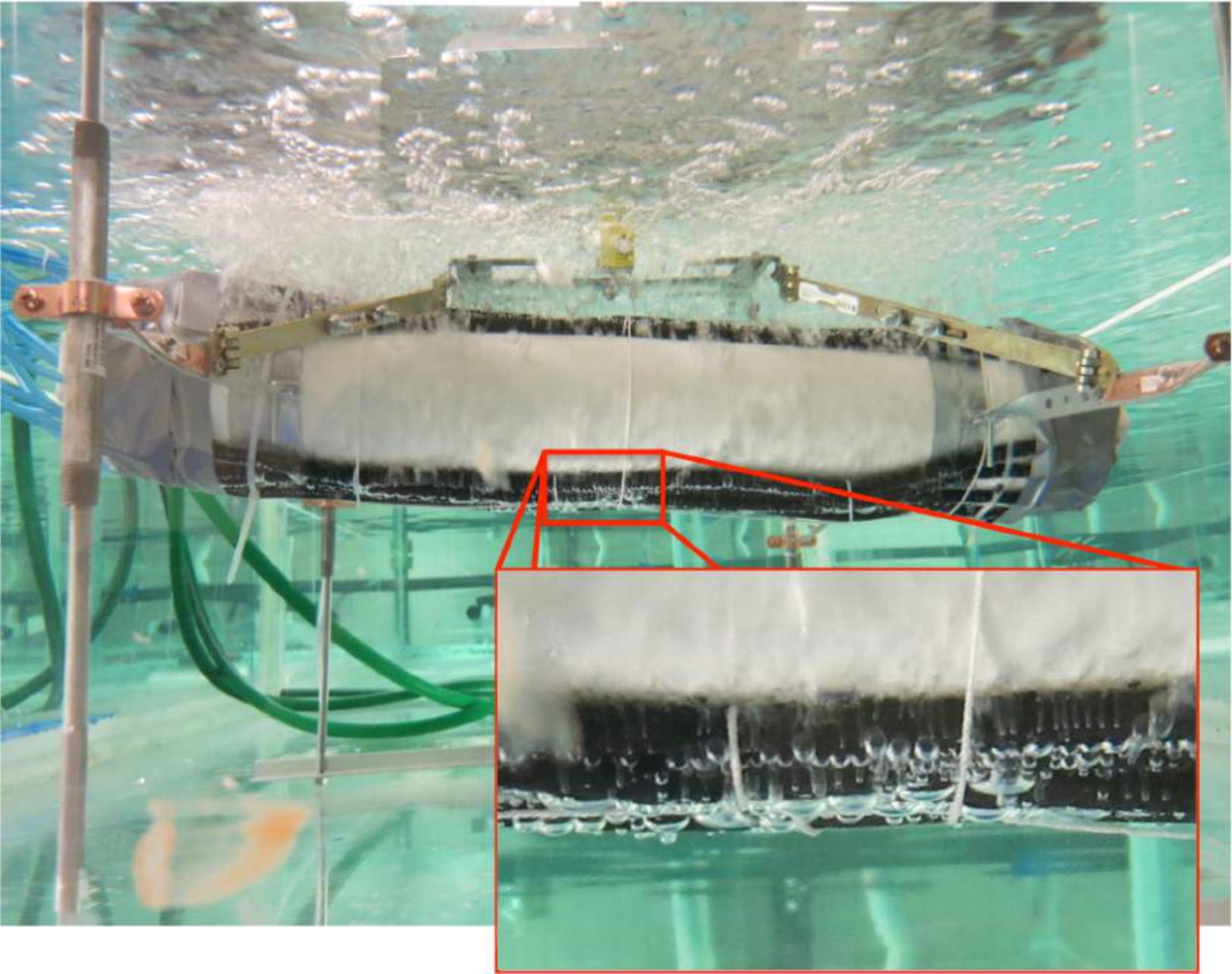}
\caption{\small{This picture is a side view of the experiment in the middle of an ongoing trial in which both the top and bottom diffusers were active. The zoomed in window offers a view of the air bubbles still attached, shortly before leaving the diffusers.}}
\label{figure 10}
\end{figure}

Consider a control volume that surrounds the test vehicle inclusive of the diffusers, such that the surface of the diffusers almost coincides with the boundary of the control volume. When the experiment was ongoing, the control volume was in a pseudo-steady state condition, thus rendering the first term on the right hand side of \eqref{p} negligible. Analyzing the control volume along the vertical axis only, the second term of the right hand side of \eqref{p} can be assumed to be zero since air bubbles left the test vehicle from the top and bottom diffusers in a nearly symmetrical fashion; thusly, their momentums cancel out to zero. All that remains is the left side of the equation. Two forces contributed to the left hand side of the equation. The first was the force exerted by the force gauge, which was read from the display and recorded in a computer. The second force is attributed to the buoyant force of the bubbles, not those already released into the surrounding water, but those still attached to the diffusers just prior to being released. Refer to the following picture for evidence Fig. \ref{figure 10}. As such, the momentum equation shows that a large part of the forces displayed by the force gauge was due to the buoyant force produced by the air bubbles before detachment from the diffusers. 

Fig. \ref{figure 10} shows that the bubbles formed were clearly not in the micro cavitation-range as initially hoped.  The next logical step was to determine how big an average air bubble must have been to be able to collectively produce a buoyant force of about $4.5~N$ taken from a scatter point of  Fig. \ref{figure 9}. A diffuser was sampled from the same batch used in this experiment and the research team manually counted an average of about 74 perforations per inch of the diffuser. Given that the diffuser was measured to be about 23.5 inches, each diffuser therefore had 1,739 perforations. A total of 16 diffusers added up to 27,824 perforations where air bubbles could form. It was assumed that the water used in this experiment was fresh water and therefore had a density of $1000~ kg/m^3$, and gravitational acceleration was $9.81 ~m/s^2$. Hence, the required volume of water displaced to produce $4.5~N$ of buoyant force must have been about $0.0004587 ~ m^3$. Assuming the bubble created at each perforation was spherical, each bubble ought to be about $3.2 ~mm$ in diameter. This appears to be a reasonable condition given that the size of the bubbles observed in Fig. \ref{figure 10} seems to conform. Note that the bubbles in reality were closer to ellipsoidal in shape, but the simplifying assumption that they were spherical was made only to obtain a rough idea of the volume of the bubbles formed. 

As explained in the previous section, due to the limitations of the compressor used, the static pressure of the compressed air dropped from about $120~psi$ at the beginning of the experiment to about $40~psi$ two minutes later. When the static pressure of the compressed air approached the lower end, the chart in Fig. \ref{figure 9} shows that the buoyant force dropped to about $ 2.7~N$. To produce such a buoyant force, each air bubble must have had an average diameter of about $2.6~mm$. Why did the size of the bubbles decrease along with the air pressure? 

An empirical equation that predicts the volume of bubble created by passing air through an orifice is \cite{Ste}:
\begin{equation}\label{eq2}
V_B=\frac{2 \pi R \sigma}{g \Delta \rho}
\end{equation}
	
\noindent where$V_B$ = bubble volume, in $cm^3$,
$R$ is orifice radius, in $cm$, $\sigma$  is surface tension of liquid, in $dynes/cm=0.1 ~Pa ~cm$, $g$ is the gravitational acceleration, in $cm/sec^2$, and
$\Delta \rho$ is difference between the density of liquid and the density of bubbles, in $g/cm^3$.

When a trial was ongoing, the orifice radius, surface tension of liquid, gravitational acceleration, and air volume were constant. However, since the air pressure dropped significantly from $120~psi$ to about $40~psi$ while its temperature did not change much (about $2\,^{\circ}{\rm C}$ during a trial duration), its density must have decreased significantly. This resulted in an increase in $\Delta \rho$ which reduced the size of the bubbles produced over time. This empirical formula is key to account for the decreasing buoyancy as a trial progressed. 

With all evidence explained, there were sufficient reasons to believe that indeed the buoyant force experienced by the underwater vessel in this experiment was a function of the number of diffusers, or more specifically, the number of air bubbles formed by the diffusers and their sizes, rather than the volume of air diffused per unit time.  This disproves the second aspect of the original hypothesis.

\section{Conclusions}

A major objective of this project was to gain a better understanding of the nature of micro cavitation bubbles, of which there is little knowledge. The result of this experiment was intended for use to gauge the effectiveness of introducing air cavities as a means of affecting an underwater vessel's motion. This technology could be very useful in practical situations such as when is embedded in the missile avoidance systems of military submarines and coast guard reconnaissance vehicles. As stated in the introduction, little was known about the way bubbles behave under water and their impact on undersea vessels, thus this experiment proved very successful at opening up new ideas and questions for further research in this vastly unexplored field. Amongst the many important lessons, two were keys in addressing the hypothesis. 

Firstly, a systematic cavitation around an undersea vessel may not be a very effective method to increase its sink rate, but it is a good way to increase its ascent rate instead. This partially proves the first part of the hypothesis, as it only accounts for the aspect where the density of the water column above a vessel decreases due to the presence of cavities, which by the Archimedes\rq{} principle it dictates that it should gain buoyancy. A slight modification in the experimental setup was needed to further investigate what would occur if the density of the overhead water column remained the same while that of the water underneath the vessel was exclusively reduced by diffusing air cavities. Perhaps, increasing the depth of the test vehicle may accentuate the effects, as there would be a greater volume of overhead water column.

\smallskip
Secondly, the rate of bubble formation and the amount of air released per unit time did not seem to affect the buoyancy of an underwater vehicle. A more important variable was the size of the bubbles formed and their quantity. This disagrees with the second facet of the hypothesis. A question arose if this result could have been influenced by the lack of control over the bubbles formed underneath the test vehicle, which was the reason why the first part of the hypothesis could not be fully proven or disproven. This possibility could be ruled out by considering the fact that the observed buoyant forces were different in the three trials even though they shared the same air volumetric flow rates. The lack of control over the bubbles formed underneath the vessel affected the density of overhead water column; but regardless of which diffusers were active, same volumetric flow rates meant that there was the same amount of air cavities in the overhead water column in all the three cases, which would translate into the same buoyant force readings. Yet, the case in which both the top and bottom diffusers were active yielded twice as much buoyant force than if only either side were active. Thus, the new conjecture asserting that the buoyant force acting on the test vehicle depends on the size of the air cavities formed and their quantity is valid. 

\smallskip
Despite the fact that the hypothesis of this experiment could not be proven with a resounding certainty, the above are very important lessons because they provide a clearer possibility of achieving control of an undersea vessel's depth using micro cavities. The top diffusers could produce bigger bubbles to fill the overhead water column with greater cavities, thus greatly enhancing the vessel's buoyancy. The bottom of the vehicle could diffuse finer cavities to lessen the buoyant effects if these bubbles were to fill up the overhead water column, yet provide enough cavities underneath to encourage the vessel to sink. To further encourage this phenomenon, the shape of the underwater vessel can be designed so that the bubbles formed underneath would not wind up in the overhead water column.

\smallskip
To achieve these, the diffuser apertures need to be varied between the top and the bottom of the underwater vehicle. The bottom apertures, especially, need to be within the micro cavitation range. The diffusers precise enough to accurately perform this experiment are almost exclusively custom made, and thus vastly more expensive than what the budget granted for this experiment could allow. 

\smallskip
Also, important to note was the lack of relationship data between air pressure and the buoyant force observed. Utilizing \eqref{eq2}, the information on air density can be obtained by measuring the pressure just upstream of the diffusers. This therefore allows for a much more precise control on the average size of the air cavities released and a better understanding of how this variable affects the buoyant force experienced by the vehicle. In addition to adding more pressure sensors, future experiments should be conducted with a larger capacity compressor that would supply a fixed and steady compressed air pressure.   Moreover, the force data recorded in this experiment could be skewed to a certain extent due to the inevitable friction between the arms of the test vehicle and the guiding steel rods. To ensure that future force readings reflect more accurate magnitudes, more than one force gauges should be used. Ideally, the guiding steel rods should be removed altogether and the terminals of the test vehicle's arms suspended directly by force gauges. Setting up the experiment as such eliminates unwanted friction entirely. Additionally, studying the differences in the force readings among the force gauges can provide more insight into the stability of the test vehicle.

\smallskip
It was briefly mentioned previously that the fluctuations in the force readings recorded in this experiment hinted at vehicle instability. This could be further investigated by examining a record of data obtained using 6-axis gyroscopes installed in future test vehicles. 
All in all, this experiment was conducted as intended, and the results were different from expectations. %Even though it did not produce conclusive results,
Important observations were noticed which could extend research opportunities.

\bigskip

\section*{ACKNOWLEDGMENT}
The authors appreciate the support from Embry-Riddle  Internal Student Research Grant FY2013-2014. 

The experiments have been conducted in the Nonlinear Waves Research Laboratory 3 (NLWL3) \cite{NLWL}

\end{document}